\documentclass[12pt,draftcls, onecolumn,journal]{IEEEtran}
\usepackage{graphicx}
\usepackage{amsmath}
\usepackage{amsfonts}
\usepackage{amssymb}
\usepackage{cite}
\usepackage{color}
\usepackage{multirow}
\usepackage{tabularx}
\newcolumntype{L}[1]{>{\raggedright\arraybackslash}p{#1}}
\newcolumntype{C}[1]{>{\centering\arraybackslash}p{#1}}
\newcolumntype{R}[1]{>{\raggedleft\arraybackslash}p{#1}}

\usepackage{times, epsfig}
\usepackage{subfig}

\newcommand{\paren}[1]{\left(#1\right)}
\newcommand{\sqparen}[1]{\left[#1\right]}
\newcommand{\brparen}[1]{\left\{#1\right\}}
\newcommand{\parenlo}[1]{\left.\left(#1\right]\right.}
\newcommand{\parenro}[1]{\left.\left[#1\right)\right.}
\newcommand{\abs}[1]{\left| #1\right|}
\newcommand{\norm}[1]{\left\| #1\right\|}
\newcommand{\field}[1]{\ensuremath{\mathbb{#1}}}
\newcommand{\N}{\ensuremath{\field{N}}} 
\newcommand{\R}{\ensuremath{\field{R}}} 
\newcommand{\C}{\ensuremath{\field{C}}} 
\newcommand{\ra}{\ensuremath{\rightarrow}} 
\newcommand{\PR}[1]{\ensuremath{\mathsf{Pr}\left\{#1\right\}}} 
\newcommand{\EW}{\ensuremath{\mathsf{E}}} 
\newcommand{\ES}[1]{\ensuremath{\mathsf{E}\left[#1 \right]}} 
\newcommand{\defeq}{\ensuremath{\triangleq}} 
\newcommand{\e}[1]{\ensuremath{{\rm e}^{#1}}} 

\newcommand{\snr}{\ensuremath{{\sf SNR}}}

\renewcommand{\vec}[1]{\ensuremath{\boldsymbol{#1}}} 
\newcommand{\pol}{\ensuremath{\mathcal{P}}}
\newcommand{\TX}{\ensuremath{\vec{x}_{\rm s}}}
\newcommand{\RX}{\ensuremath{\vec{x}_{\rm d}}}
\newcommand{\rateins}{\ensuremath{{\sf r}_{\varphi}}}
\newcommand{\rateave}{\ensuremath{{\sf R}_{\Phi}}}
\newcommand{\Rave}{\ensuremath{{\sf R}_{\rm ave}}}
\newcommand{\xopt}{\ensuremath{\vec{x}_{\rm opt}}}

\newcommand{\xmid}{\ensuremath{\vec{x}_{\rm mid}}}

\newcommand{\xs}{\ensuremath{\vec{x}_{\rm s}}}
\newcommand{\xd}{\ensuremath{\vec{x}_{\rm d}}}

\newcommand{\Hsd}{\ensuremath{H_{\rm s, d}}}
\newcommand{\Hsr}{\ensuremath{H_{\rm s, r}}}
\newcommand{\Hrd}{\ensuremath{H_{\rm r, d}}}
\newcommand{\Pout}{\ensuremath{{\sf P}_{\rm out}}}
\newcommand{\relayselect}{\ensuremath{\widehat{s}_{}}}

\DeclareMathOperator{\arcsec}{arcsec}

\DeclareMathOperator{\arccsc}{arccsc}

\IEEEoverridecommandlockouts

\newtheorem{definition}{Definition}

\newtheorem{theorem}{Theorem}

\newtheorem{lemma}[theorem]{Lemma}

\newlength{\figwidth}
\begin{document}

\setlength{\pdfpagewidth}{8.5in}
\setlength{\pdfpageheight}{11in}

\title{Location-Based Optimum Relay Selection in \\ Random Spatial Networks}
\vspace{0mm}

\author{
\IEEEauthorblockN{Saman Atapattu, Hazer Inaltekin  and Jamie Evans\\}
\thanks{S.~Atapattu  and J.~Evans are with the Department of Electrical and Electronic Engineering, University of Melbourne, Parkville, VIC 3010, Australia.}
\thanks{H.~Inaltekin is with the School of Engineering, Macquarie University, North Ryde, NSW 2109, Australia.}
}

\maketitle

\begin{abstract}
This paper investigates the location-based relay selection problem, where the source node chooses its relay from a set of spatially deployed decode-and-forward relays. The advantages of location-based relay selection are the elimination of excessive relay switching rate and the feedback reduction avoiding the requirement of having full channel state information at the source node. For a homogeneous  Poisson point process of candidate relays, we first derive the distribution for the distance of the relay (relative to the source and destination nodes) selected by the {\em optimum} location-based relay selection policy. This result is independent of the functional form of the path-loss function as long as it is a non-increasing function of the transmitter-receiver separation. By utilizing the derived optimum relay distance distribution, we then obtain analytical expressions for the average rate and outage probability by considering the power-law decaying path-loss function for the no-fading and Rayleigh fading communication scenarios. It is observed that the optimum relay selection policy outperforms the other common selection strategies notably, including the ones choosing the relay closest to the source, the relay closest to the destination and the relay closest to the mid-point between source and destination. 
\end{abstract}

\newpage
\section{Introduction}
The fifth generation (5G) of wireless networks are expected to provide high data rates by using larger bandwidths. While the millimeter wave (mmWave) frequency bands offer
this opportunity, the direct communication suffers from higher propagation losses 
due to blockages \cite{Yanikomeroglu04}. To mitigate such drawbacks, cooperative relay nodes  can be used to enhance coverage, rate and reliability \cite{Iwamura10,Yuan13,Atapattu13,Fan2018acc}. 
Although using multiple relays for a single-communication link can improve the performance, overall system complexity and signaling overhead also increase due to synchronization issues and signal processing cost for combining.  
Therefore, activating only one or a few relays from a large set of relay nodes is a practical approach 
for relay-aided wireless networks design to reduce excessive resource utilization and overhead \cite{Atapattu2018tcom,Atapattu2019twc}.

There has been some previous work for relay selection using spatial network models, e.g., see \cite{Cho2011tvt, Galarza14, Tukmanov2014tcom, Zhou2016twc, Elkotby2015gcom, Krikidis2014tcom} and references therein. In \cite{Cho2011tvt}, the authors focused on reducing relay selection signaling overhead by characterizing a quality-of-service (QoS) region satisfying the target outage probability. Without considering any time-scale separation between fading and relay location processes, they also studied the outage performance of both random and best relay selections from this QoS region.   
In \cite{Galarza14}, the optimum relay activation probability was analyzed for decode-and-forward (DF) relays selected based on their distances to the source node. 
In \cite{Tukmanov2014tcom}, the outage probability was analyzed with imperfect channel state information (CSI) when the relay having the best channel to the destination is selected to forward messages. 
A similar relay selection was also considered in \cite{Zhou2016twc}, deriving the outage probability and average packet delay. 
In \cite{Elkotby2015gcom}, a random relay selection policy choosing a relay in an area around the source-destination mid-point 
was proposed and its outage performance was compared with the nearest-neighbor relay selection. 
In \cite{Krikidis2014tcom}, a random relay selection process selecting a single relay out of all potential relays in a feasible region was considered for an energy harvesting network. 

In this paper, different from the above previous work, we investigate the performance of the {\em optimum} relay selection policy for general path-loss models when there is no direct link between source and destination nodes. We consider the time-scale separation between fading and relay location processes cautiously, and identify its effect on the network performance. Our selection criterion is based only on relay locations for selecting the relay node. This approach is reminiscent of the base-station selection strategy in cellular networks, and the optimum one when only location information but not the full CSI is available at the source node.\footnote{Fading usually changes at a much faster time scale than the node locations \cite{Tse05}. Hence, it is not practical for the source node to obtain CSI for all source-to-relay and relay-to-destination channels, and establish a connection to another relay node for handing over the data traffic each time fading coefficients change. This approach will result in excessive signaling overhead and relay switching rate.} For the optimum relay selection policy, 
we first derive the optimum relay distance distribution. Then, by utilizing the derived distribution,  the average rate and outage probability are evaluated for both no-fading and Rayleigh fading scenarios. Finally, the derived analytical expressions are verified and the performance improvement achieved by the optimum relay selection policy is illustrated through extensive simulations. 

{\it Notation}: We use boldface, upper-case and calligraphic letters to represent vector quantities, random variables and sets, respectively. We use $\R$, $\N$, $\C$ and $\R^2$ to denote the real, natural and complex numbers, and the two-dimensional Euclidean space, respectively. $| x |$ denotes the absolute value of a scalar quantity $x$ (real or complex), whereas $\| \vec{x} \|$ is used to measure the canonical Euclidean norm of a vector quantity $\vec{x}$. 
Expected value of a random variable $X$ is denoted by $\ES{X}$.

\section{Network Model and Performance Metrics}\label{s_sys}


\subsection{Network Model} \label{Subsection: Network Model} 
\begin{figure}[!t]
\begin{center}
\includegraphics[scale=0.7]{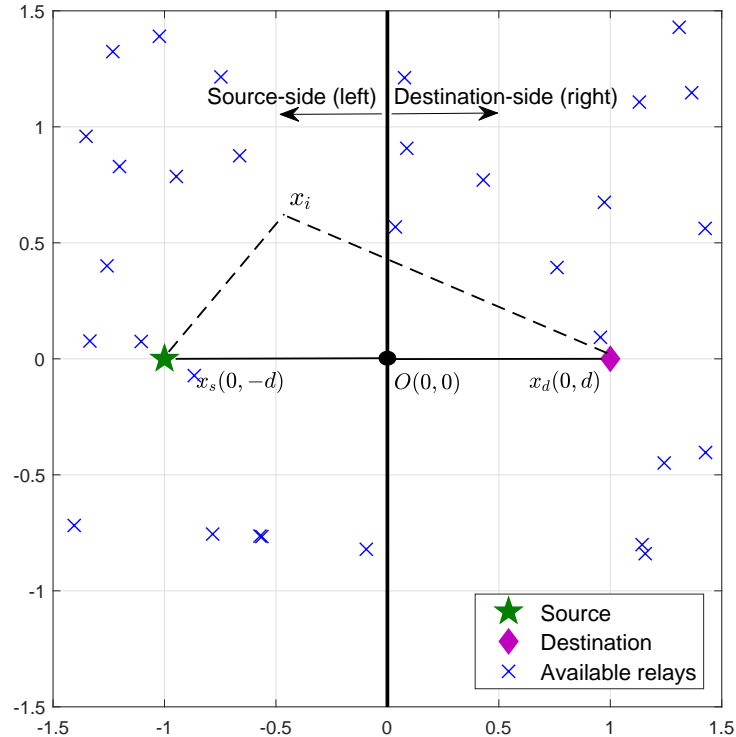}
\end{center}
\caption{An example configuration for a relay-aided wireless network.} \label{Fig: System Model} 
\end{figure}   
We consider a relay-aided {\em spatial} wireless network in $\R^2$, as illustrated by Fig. \ref{Fig: System Model}. The network contains a source-destination pair having arbitrary locations $\TX \in \R^2$ (source node) and $\RX \in \R^2$ (destination node). The locations of potential DF relay nodes in the network are given by $\varphi = \brparen{\vec{x}_1, \vec{x}_2, \ldots}$, where $\vec{x}_i \in \R^2$ represents the $i$th DF relay location for $i \in \N$.  We will always assume that $\varphi$ is a locally finite set, i.e., there are only finitely many relays in every bounded subset of $\R^2$. For performance analysis, we will consider the communication scenario in which relay locations are random and determined according to a spatial homogeneous Poisson point process (HPPP) $\Phi = \brparen{\vec{X}_1, \vec{X}_2, \ldots }$. Hence, $\varphi$ should be interpreted as a particular realization of $\Phi$ below. 
We assume that the primary aim of the relays is to assist data communication between source and destination nodes without generating any additional traffic, as in the latest proposals for LTE-A standards \cite{3GPPRelay2, 3GPPRelay1}.  
%
We examine the network performance under a given relay selection policy, which is formally defined as follows.      
\begin{definition} \label{Def: Selection Policy}
A relay selection policy $\pol: \Sigma \mapsto \R^2$ is a mapping from the set of all countable locally finite subsets of $\R^2$, denoted by $\Sigma$, to $\R^2$ satisfying the condition $\pol\paren{\varphi} \in \varphi$ for all $\varphi \in \Sigma$.    
\end{definition}

For the sake of notational simplicity, we parametrize the relay location selected by $\pol$ as $\vec{x}_\pol$ in the remainder of the paper, with the understanding that $\vec{x}_\pol = \pol\paren{\varphi}$ for any given $\varphi \in \Sigma$. The same meaning is attributed to other variables throughout the paper when we use this parametrization for them as well. Then, the {\em instantaneous} rate of information flow from source to destination (in bits/sec/Hz) as a function of the relay selection policy $\pol$ is given as 
\begin{eqnarray}
&\rateins\paren{\pol} = & \frac12 \min\biggl\{\log_2\paren{1 + \snr \abs{\Hsr}^2G\paren{\norm{\TX - \vec{x}_\pol}}},\nonumber \\& & \log_2\paren{1 + \snr\paren{\abs{\Hsd}^2G\paren{\norm{\TX - \RX}} + \abs{\Hrd}^2G\paren{\norm{\vec{x}_\pol - \RX}}}} \biggr\} \label{Eqn: DF Rate 1}
\end{eqnarray}
%
%
where $G$ is a non-increasing path-loss function, $H_{a,b} \in \C$ is the random fading coefficient between the source, {\em selected} relay and destination nodes for $a \in \brparen{{\rm s, r}}$ and $b \in \brparen{{\rm r, d}}$, 
and $\snr \defeq \frac{P}{W N_0}$ is the signal-to-noise ratio with $P$, $W$ and $N_0$ being transmission power, communication bandwidth and noise energy per complex dimension, respectively, \cite{Gallager08}.  The rate $\rateins\paren{\pol}$ in \eqref{Eqn: DF Rate 1} is achievable for each fading state $\vec{H} = \paren{\Hsr, \Hrd, \Hsd}^\top$ by using independent Gaussian codebooks at the source and relay nodes and CSI at the receivers \cite{Laneman04}.\footnote{Implicit in this formulation is the assumption of a perfect multiple-access and interference cancellation scheme so that interference is only seen as the background noise at the receivers.}     
We will assume in the remainder of the paper that the direct link between source and destination nodes is severely shadowed by an object in the environment (i.e., $\abs{\Hsd} \approx 0$), or the sharp decay rate of $G$ with distance as in the mmWave communications \cite{Iwamura10,Yuan13,Yanikomeroglu04}. 
%
We will also assume that random fading coefficients $\Hsr$ and $\Hrd$ change at a much faster time-scale than the network node locations. 
Hence, our selection criterion will be based only on relay locations for selecting the relay node, which is also embodied in Definition \ref{Def: Selection Policy}.  
\vspace{-1mm}
\subsection{Performance Metrics} \label{Subsection: Performance Metrics} 
For determining the performance of a relay selection policy $\pol$, we will use the average data rate and outage probability as our performance metrics \cite{Tse05}. 
We assume for both cases that the permissible decoding delay is large enough to average over the fading process. Then, given $\pol$ and the random relay locations $\Phi = \brparen{\vec{X}_1, \vec{X}_2, \ldots}$, the rate \[\EW\sqparen{\log_2\paren{1 + \snr \abs{\Hsr}^2G\paren{\norm{\TX - \vec{X}_\pol}}} \big| \Phi}\]
is achievable over the wireless link connecting the source and selected relay, where the expectation is taken over the randomness due to fading. Similarly, the rate 
\[\EW\sqparen{\log_2\paren{1 + \snr \abs{\Hrd}^2G\paren{\norm{\vec{X}_\pol - \RX}}} \big| \Phi}\]
is achievable between the selected relay and destination. 
Hence, we can express the  data rate, averaged over the fading process, from source to destination according to
\begin{eqnarray}
&\rateave\paren{\pol} = & \frac12 \min\biggl\{\EW\sqparen{\log_2\paren{1 + \snr \abs{\Hsr}^2G\paren{\norm{\TX - \vec{X}_\pol}}} \big| \Phi}, \nonumber\\ & & \EW\sqparen{\log_2\paren{1 + \snr \abs{\Hrd}^2G\paren{\norm{\vec{X}_\pol - \RX}}} \big| \Phi} \biggr\} \label{Eqn: Rate Given Locations}
\end{eqnarray}
which is a function of $\pol$ and $\Phi$. Using $\rateave\paren{\pol}$ in \eqref{Eqn: Rate Given Locations}, the outage probability for delay-sensitive traffic is defined as below. 


\begin{definition} \label{Def: Outage Probability}
For a target bit rate $\rho$ and a relay selection policy $\pol$, the outage probability $\Pout\paren{\pol}$ is equal to 
\begin{eqnarray}
\Pout\paren{\pol} = \PR{\rateave\paren{\pol} \leq \rho}. \label{Eqn: Outage Probability}
\end{eqnarray}
\end{definition}

Similarly, we define the average rate when the decoding delay can be relaxed to permit variable rate transmission as follows.  
\begin{definition} \label{Def: Average Rate}
For a given relay selection policy $\pol$, the average rate is defined to be 
\begin{eqnarray}
\Rave\paren{\pol} = \ES{\rateave\paren{\pol}}, \label{Eqn: Average Rate}
\end{eqnarray}
where the expectation 
is over the random relay locations. 
\end{definition}

It is important to note that $\Rave\paren{\pol}$ defined in \eqref{Eqn: Average Rate} should be interpreted as the average of the rate in \eqref{Eqn: Rate Given Locations}, which is averaged over the random 
relay locations. It is not the ergodic rate achieved over the long time-horizon for the same source, relay and destination nodes. Rather, the rate in \eqref{Eqn: Rate Given Locations} is attained with a different relay node selected by $\pol$ for each realization of $\Phi$.  
In addition, we observe that we can change the order of minimum and expectation operators while going from \eqref{Eqn: DF Rate 1} to \eqref{Eqn: Rate Given Locations}. This is because the decoding delays permit averaging over the fading process before switching to another relay node. It is more advantageous to minimize expected rates than averaging the minimum of instantaneous rates. This is the {\em waiting-gain} we have 
in \eqref{Eqn: Rate Given Locations}. 

\section{Optimum Relay Selection Problem} \label{Section: Optimum Relay Selection Problem}
We start with the goal of maximizing $\rateins\paren{\pol}$ by optimizing $\pol$ for each $\varphi \in \Sigma$ without knowledge of fading states at the source node. The solution of this optimization problem is also the one optimizing the performance metrics $\Pout\paren{\pol}$ and $\Rave\paren{\pol}$, as discussed below. 
Using the monotonicity of the path-loss function, the  problem of maximizing $\rateins\paren{\pol}$ is equivalent to minimizing the relay selection function $\relayselect\paren{\vec{x}}$, which is given by
\begin{eqnarray}
\relayselect\paren{\vec{x}} = \max\brparen{\norm{\xs - \vec{x}}, \norm{\vec{x} - \xd}}, \label{Eqn: Relay Selection Function}
\end{eqnarray}
over the set of relay locations in $\varphi$.  Hence, the optimum relay selection problem can be written as the following discrete optimization problem:
\begin{eqnarray}
\begin{array}{ll}
\underset{\vec{x} \in \R^2}{\mbox{minimize}} & \relayselect\paren{\vec{x}} \\
\mbox{subject to} & \vec{x} \in \varphi
\end{array} \label{Eqn: Relay Selection Problem}
\end{eqnarray}
for each $\varphi \in \Sigma$. The solution of \eqref{Eqn: Relay Selection Problem} is well-defined and belongs to $\varphi$ for all $\varphi \in \Sigma$ since $\varphi$ is assumed to be a locally finite countable set. The optimum relay selection policy, which we denote by $\pol_{\rm opt}$, is the one that solves \eqref{Eqn: Relay Selection Problem} for all $\varphi \in \Sigma$. When we write $\xopt$, we will refer to the location of the relay selected by $\pol_{\rm opt}$. Observing the structure of $\relayselect\paren{\vec{x}}$ and min-max type of optimization in \eqref{Eqn: Relay Selection Problem}, it is easy to see that the optimum relay location in $\varphi$ must have the important property of balancing distances to both source and destination nodes. Therefore, an efficient relay selection policy should also have such a distance balancing property. 
We will investigate this balancing property in detail for the mid-point relay selection policy and obtain its performance for HPPPs in Section \ref{Section: Mid-Point Relay Selection}.  

The connection between the solution of the optimization problem in \eqref{Eqn: Relay Selection Problem} and the performance metrics $\Pout\paren{\pol}$ and $\Rave\paren{\pol}$ is not explicit.  In Lemma \ref{Lemma: Rate and Outage Optimality} below, we relate the solution of \eqref{Eqn: Relay Selection Problem} to $\Pout\paren{\pol}$ and $\Rave\paren{\pol}$ by showing that the optimum relay selection policy solving \eqref{Eqn: Relay Selection Problem} for each $\varphi \in \Sigma$ is also the best choice for optimizing $\Pout\paren{\pol}$ and $\Rave\paren{\pol}$. 
\begin{lemma} \label{Lemma: Rate and Outage Optimality}
Let $\Xi$ be the set of all feasible relay selection policies. Then, for identically distributed fading at each wireless link, we have
\begin{eqnarray}
\Pout\paren{\pol_{\rm opt}} = \inf_{\pol \in \Xi} \Pout\paren{\pol} \label{Eqn: Outage Probability Optimality}
\end{eqnarray}
\begin{eqnarray}
\Rave\paren{\pol_{\rm opt}} = \sup_{\pol \in \Xi} \Rave\paren{\pol}. \label{Eqn: Average Rate Optimality}
\end{eqnarray}
\end{lemma}
\begin{IEEEproof}
We first observe the following trivial inequality. If $X_1$ and $X_2$ are two identically distributed non-negative random variables, and $c_1$ and $c_2$ are two non-negative arbitrary constants satisfying $c_1\leq c_2$, then $\ES{\log_2\paren{1 + c_1 X_1}} \leq \ES{\log_2\paren{1 + c_2 X_2}}$. Using this inequality and recalling that for any given relay selection policy $\pol$, $\vec{X}_\pol$ is the location of the relay node selected by $\pol$ when it runs over $\Phi$, we can write
\begin{eqnarray} 
\rateave\paren{\pol} = \frac12 \EW\sqparen{\log_2\paren{1 + \snr \abs{H}^2G\paren{\relayselect\paren{\vec{X}_\pol}}} \big| \Phi}  \nonumber
\end{eqnarray}
where $H$ is the generic fading random variable having the same distribution with $\Hsr$ and $\Hrd$. Since $\pol_{\rm opt}$ solves \eqref{Eqn: Relay Selection Problem} for each $\varphi \in \Sigma$, we also have $\relayselect\paren{\vec{X}_{{\rm opt}}} \leq \relayselect\paren{\vec{X}_\pol}$. Using the above inequality one more time and the property that $G$ is a non-increasing function, we conclude that $\rateave\paren{\pol_{\rm opt}} \geq \rateave\paren{\pol}$ for any $\pol \in \Xi$. This implies $\Rave\paren{\pol_{\rm opt}} = \sup_{\pol \in \Xi} \Rave\paren{\pol}$ and $\Pout\paren{\pol_{\rm opt}} = \inf_{\pol \in \Xi} \Pout\paren{\pol}$.   
\end{IEEEproof}



\section{Optimum Relay Selection and Its Performance} \label{Section: Optimum Relay Selection}
In this section, we will obtain analytical expressions for $\Rave\paren{\pol_{\rm opt}}$ and $\Pout\paren{\pol_{\rm opt}}$. To this end, we will take random relay location process $\Phi$ as an HPPP having intensity $\lambda>0$. Due to the stationarity property of HPPPs \cite{Kingman93}, we will assume that $\xs = \paren{-d, 0}^\top$ and $\xd = \paren{0, d}^\top$ without loss of generality. This is the example network configuration illustrated in Fig. \ref{Fig: System Model}.    

\subsection{The Optimum Relay Distance Distribution} \label{Section: Optimum Relay Distance}
In the proof of Lemma \ref{Lemma: Rate and Outage Optimality}, we showed that the conditional data rate, given the relay locations, is equal to 
\begin{eqnarray}\label{Eqn: Conditional Average Rate}
\rateave\paren{\pol} = \frac12 \EW\sqparen{\log_2\paren{1 + \snr \abs{H}^2G\paren{\relayselect\paren{\vec{X}_\pol}}} \big| \Phi}
\end{eqnarray}
for any relay selection policy. Hence, the key step to obtain analytical expressions for $\Rave\paren{\pol_{\rm opt}}$ and $\Pout\paren{\pol_{\rm opt}}$ is to derive the distribution function for $\relayselect\paren{\vec{X}_{{\rm opt}}}$. For notational simplicity, we define $\Gamma_{\rm opt} \defeq \relayselect\paren{\vec{X}_{{\rm opt}}}$. Then, by definition of $\pol_{\rm opt}$, we have
\begin{eqnarray}
\Gamma_{\rm opt} = \min_{\vec{X} \in \Phi} \relayselect\paren{\vec{X}}. \label{Eqn: Optimum Relay Distance} 
\end{eqnarray}
That is, $\Gamma_{\rm opt}$ is the minimum value achieved by the relay selection function over the HPPP $\Phi$. In the next theorem, we provide the cumulative distribution function (cdf) and the probability distribution function (pdf) of $\Gamma_{\rm opt}$. 

\begin{theorem} \label{Theorem: Optimal dis distribution}
The cdf of $\Gamma_{\rm opt}$ is given by 
\begin{equation}\label{e_cdfmaxminD}
\begin{split}
\hspace{-2mm}F_{\Gamma_{\rm opt}}\paren{\gamma} = \left\{
\begin{array}{ll}
\hspace{-2mm}0 & 
\gamma < d \\
\hspace{-2mm}1-\e{2\lambda\paren{d\sqrt{\gamma^2-d^2}-\gamma^2 \arcsec\paren{\frac{\gamma}{d}}}} & 
\gamma \geq d
\end{array}. \right.
\end{split}
\end{equation}
Similarly, the pdf of $\Gamma_{\rm opt}$ is given by for $\gamma \geq d$
\begin{equation}\label{e_pdfmaxminD}
\begin{split}
\hspace{-2mm}f_{\Gamma_{\rm opt}}\paren{\gamma} & = 4 \lambda  \gamma \arcsec\left(\frac{\gamma}{d}\right) \e{2\lambda  \left(d \sqrt{\gamma^2-d^2}-\gamma^2 \arcsec\left(\frac{\gamma}{d}\right)\right)}.
\end{split}
\end{equation}
\end{theorem}
\begin{IEEEproof}
See Appendix \ref{Appendix: optimal dis distribution Proof}.
\end{IEEEproof}


Our results presented up to this point is independent of the functional form of the path-loss function as long as it is a non-increasing function of the transmitter-receiver separation. However, for obtaining numerical performance curves, we will focus on the classical power-law decaying path-loss function $G\paren{x} = \frac{1}{x^\alpha}$, where $\alpha \geq 2$ is the path-loss exponent. Hence, we will frequently use the distribution of $\frac{\snr}{\Gamma_{\rm opt}^\alpha}$ in the rest of the paper. We provide its cdf and pdf of $\frac{\snr}{\Gamma_{\rm opt}^\alpha}$ in the following lemma for the sake of presentation clarity. 
\begin{lemma} \label{Lemma: Path loss dis}
Let $Y = \frac{\snr}{\Gamma_{\rm opt}^\alpha}$ for a path-loss exponent $\alpha\geq 2$. Then, the cdf and pdf of $Y$ are, respectively,  equal to 
\begin{equation}\label{e_cdfmaxminDalp}
\begin{split}
F_{Y}(y)  = \e{2 \lambda  \left(d \sqrt{\left(\frac{\snr}{y}\right)^{\frac{2}{\alpha} }-d^2}-\left(\frac{\snr}{y}\right)^{\frac{2}{\alpha}} \arcsec\left(\frac{1}{d}\left(\frac{\snr}{y}\right)^{\frac{1}{\alpha}}\right)\right)}
\end{split}
\end{equation}
and
\begin{equation}\label{e_pdfmaxminDalp}
\begin{split}
f_{Y}(y)  = \frac{4 \lambda}{\alpha  y}  \left(\frac{\snr}{y}\right)^{\frac{2}{\alpha}} \arcsec\left(\frac{1}{d}\left(\frac{\snr}{y}\right)^{\frac{1}{\alpha}}\right)  \e{2 \lambda  \left(d \sqrt{\left(\frac{\snr}{y}\right)^{\frac{2}{\alpha} }-d^2}-\left(\frac{\snr}{y}\right)^{\frac{2}{\alpha}} \arcsec\left(\frac{1}{d}\left(\frac{\snr}{y}\right)^{\frac{1}{\alpha}}\right)\right)}
\end{split}
\end{equation} 
if $y\leq \frac{\snr}{d^\alpha}$, and zero otherwise. 

\end{lemma}
\begin{IEEEproof}
Let $g$ be a mapping given by $g(x) = \frac{\snr}{x^\alpha}$. Then, $Y = g\paren{\Gamma_{\rm opt}}$. Since $g$ is a strictly decreasing function of its argument for $x > 0$, the cdf and pdf of $Y$ can be written as $F_Y(y) = 1 - F_{\Gamma_{\rm opt}}\paren{g^{-1}(y)}$ and $f_Y(y) = -f_{\Gamma_{\rm opt}}\paren{g^{-1}(y)}\frac{d}{dy}g^{-1}(y)$ for all $y \geq \frac{\snr}{d^\alpha}$ with the aid of variable transformation. For $y < \frac{\snr}{d^\alpha}$, $F_Y(y)=0$ and $f_Y(y)=0$ because $Y$ always takes values greater than or equal to $\frac{\snr}{d^\alpha}$. Calculating the inverse $g^{-1}$ and using \eqref{e_cdfmaxminD} and \eqref{e_pdfmaxminD}, we obtain $F_{Y}(y)$ and $f_{Y}(y)$ given in \eqref{e_cdfmaxminDalp} and \eqref{e_pdfmaxminDalp}.         
\end{IEEEproof}
\vspace{0mm}
\subsection{Average Rate}\label{ss_tp}
Now, we derive analytical expressions for the average rate achieved by the optimum relay selection policy $\pol_{\rm opt}$. We will consider both no-fading and Rayleigh fading cases. For the no-fading case, we assume a deterministic normalized channel gain with unit power, i.e., $\abs{H} = 1$.  For the Rayleigh fading case, we consider that $H$ is a circularly symmetric complex Gaussian random variable with unit power, i.e., $H \sim \mathcal{CN}(0,1)$. Hence, $\abs{H}^2$ is an exponential random variable with unit mean, i.e., $f_{\abs{H}^2}(x) = e^{-x}$. For the path-loss model, we consider the power-law decaying path-loss function $G(x) = \frac{1}{x^\alpha}$ for $\alpha \geq 2$.  
Using \eqref{Eqn: Conditional Average Rate} and \eqref{Eqn: Optimum Relay Distance}, we have  
\begin{equation}\label{throughput given phi}
\begin{split}
\rateave\paren{\pol_{\rm opt}}  = \frac12 \EW\sqparen{\log_2\paren{1 + \frac{\snr \abs{H}^2}{\Gamma_{\rm opt}^\alpha}} \Big| \Phi} 
=\left\{
\begin{array}{ll}
\hspace{-2mm}\frac{1}{2}\log_2\left(1+\frac{\snr }{\Gamma_{\rm opt}^\alpha}\right) &\text{if no-fading}\\
\hspace{-2mm}\frac{1}{2\ln2} e^{\frac{\Gamma_{\rm opt}^\alpha}{\snr}} \text{E}_1\left(\frac{\Gamma_{\rm opt}^\alpha}{\snr}\right) &\text{if Rayleigh}
\end{array},\right.
\end{split}
\end{equation}
where the last equality follows from the identity $b\int_{0}^{\infty}\log(1+ax)\e{-bx}dx=\e{\frac{b}{a}}\text{E}_1\left(\frac{b}{a} \right)$, and $\text{E}_1\left(\cdot \right)$ is the exponential integral \cite[eq.~8.211.1]{Gradshteyn07_book}.  

From {Definition \ref{Def: Average Rate}}, the average rate, averaged over $\Phi$, can be calculated  as 
\begin{equation}\label{e_avgtp}
\begin{split}
\Rave\paren{\pol_{\rm opt}}  = \left\{
\begin{array}{ll}
\hspace{-2mm}\frac{1}{2}\int_{0}^{\frac{\snr}{d^\alpha}}\log_2\left(1+y\right)f_{Y}(y)dy  &\text{if no-fading}\\
\hspace{-2mm}\frac{1}{2\ln2}\int_0^{\frac{\snr}{d^\alpha}} e^{\frac{1}{y}} \text{E}_1\left(\frac{1}{y}\right)f_{Y}(y) dy &\text{if Rayleigh}
\end{array}. \right.
\end{split}
\end{equation}
It is not possible to reduce the integral expression in \eqref{e_avgtp} for $\Rave\paren{\pol_{\rm opt}}$ to a closed-form. However, this single integral can be evaluated very quickly and efficiently through standard numerical integration techniques. 

\subsection{Outage Probability}\label{ss_out}
The rate outage probability is the probability that the rate  falls below a certain predetermined threshold $\rho$. From {Definition~\ref{Def: Outage Probability}}, we write
$\Pout\paren{\pol_{\rm opt}} = \PR{\rateave\paren{\pol_{\rm opt}} \leq \rho}$.
For the no-fading case, with the aid of \eqref{e_cdfmaxminDalp} and \eqref{throughput given phi}, we have
\begin{equation}\label{e_pout_nofad}
\begin{split}
\hspace{-2mm} \Pout\paren{\pol_{\rm opt}}  =\left\{
\begin{array}{ll}
\hspace{-2mm}F_Y\left( 2^{2\rho} -1 \right)& \hspace{-2mm} \mbox{if } \rho \leq\frac 1 2 \log_2\left(1+\frac{\snr}{d^\alpha}\right)\\
\hspace{-2mm}1& \hspace{-2mm}\text{if } \rho > \frac 1 2 \log_2\left(1+\frac{\snr}{d^\alpha}\right)
\end{array}.\right.
\end{split}
\end{equation}
Similarly, for the Rayleigh fading case, we have
\begin{equation}\label{e_pout_ray}
\begin{split}
\hspace{-1mm} \Pout\paren{\pol_{\rm opt}} 
  & = \left\{
\begin{array}{ll}
\hspace{-2mm}F_Y\left( y^\star \right) & \mbox{if } y^\star \leq\frac{\snr}{d^\alpha}\\
\hspace{-2mm}1 &\text{if } y^\star > \frac{\snr}{d^\alpha}
\end{array},\right.
\end{split}
\end{equation}
where $y^\star$ is the {\em unique} solution for the equation
$e^{1/y} \text{E}_1\left(1/y\right) = 2\rho\ln2.$
Uniqueness of $y^\star$ follows from the fact that $e^{1/y} \text{E}_1\left(1/y\right)$ increases monotonically with $y$. Hence, it can be readily calculated by using numerical methods such as the bisection technique or by using Matlab's \texttt{vpasolve} routine. 
Moreover, by using the inequality \cite[p.~229,~5.1.20]{Abramowitz1974} 
we can also bound $\Pout\paren{\pol_{\rm opt}}$ for Rayleigh fading as 
$F_Y\left(2^{2\rho} -1 \right) < \Pout\paren{\pol_{\rm opt}} < F_Y\left( \frac{2^{4\rho} -1}{2} \right)$. 

\section{Mid-point Relay Selection Policy} \label{Section: Mid-Point Relay Selection}

For comparison purposes, we now consider a mid-point relay selection policy $\pol_{\rm mid}$, which selects the relay closet to the mid-point between source and destination nodes. We will denote the mid-point between source and destination nodes as $\vec{w}$ i.e., $\vec{w} = \frac{\xs + \xd}{2}$.  The following is a straightforward result formally stating the distance balancing property mentioned in Section \ref{Section: Optimum Relay Selection Problem}, which we provide for the sake of completeness. 
\begin{lemma} \label{Lemma: Mid-Point Property}
For all $\vec{x} \in \R^2$, $\relayselect\paren{\vec{w}} \leq \relayselect\paren{\vec{x}}$.
\end{lemma}
\begin{IEEEproof}
By triangle inequality, we have $
\norm{\xs - \vec{x}} + \norm{\vec{x} - \xd} \geq 2d
$
for all $\vec{x} \in \R^2$, which implies that  $\relayselect\paren{\vec{x}} \geq d$ for all $\vec{x} \in \R^2$. This lower is achieved with equality for $\vec{w}$. 
\end{IEEEproof}

It should be first noted that $\pol_{\rm mid}$ is not always optimum since, in addition to its distance from $\vec{w}$, 
the relay's orientation is also crucial in determining the value of $\relayselect\paren{\vec{x}}$. That is, the closer the relay location $\vec{x}$ to the equidistant hyperplane $\mathcal{H}$ 
between $\xs$ and $\xd$ is, the smaller the value that $\relayselect\paren{\vec{x}}$ takes. This simple geometrical fact will be helpful to obtain a sufficient condition under which $\pol_{\rm mid}$ is optimal. 
Hence, we formally state it in the following lemma for the sake of completeness again.     

\begin{lemma} \label{Lemma: Orientation Property}
Let 
\[\mathcal{H} = \brparen{\vec{x} \in \R^2: \paren{\xs - \xd}^\top \vec{x} = \frac12\paren{\norm{\xs}^2 - \norm{\xd}^2}}\] 
be the equidistant hyperplane between $\xs$ and $\xd$. 
For a given $\vec{y} \in \R^2$, let also 
\[\mathcal{C} = \brparen{\vec{x} \in \R^2: \norm{\vec{x} - \vec{w}} = \norm{\vec{y} - \vec{w}}},\] i.e., $\mathcal{C}$ is the circle around $\vec{w}$ with radius $\norm{\vec{y} - \vec{w}}$. Then, for $\vec{x} \in \mathcal{H} \cap \mathcal{C}$, we have 
$\relayselect\paren{\vec{x}} = \sqrt{d^2 + \norm{\vec{y} - \vec{w}}^2}$ and $\relayselect\paren{\vec{x}} \leq \relayselect\paren{\vec{y}}$. 
\end{lemma}
\begin{IEEEproof}
See Appendix \ref{Appendix: Orientation Property Proof}.
\end{IEEEproof}

Next, using Lemma \ref{Lemma: Orientation Property}, we obtain a sufficient condition for the optimality of $\pol_{\rm mid}$ in the next theorem. 
\begin{theorem} \label{Lemma: Mid-point Optimality}
Let 
$\xmid \in \varphi$ be the relay location selected by $\pol_{\rm mid}$ and $\vec{x}_{(2)} \in \varphi$ be the location of the second closest relay to $\vec{w}$. Then, $\pol_{\rm mid}$ solves \eqref{Eqn: Relay Selection Problem}, i.e., $\relayselect\paren{\xmid} = \relayselect\paren{\xopt}$, if $\relayselect\paren{\xmid} \leq \sqrt{d^2 + \norm{\vec{x}_{(2)} - \vec{w}}^2}$.   
\end{theorem}
\begin{IEEEproof}
Let $\vec{x}_{(1)}, \vec{x}_{(2)}, \ldots$ be the ordering of relay locations in $\varphi$ in an ascending manner with respect to their distances to $\vec{w}$, i.e., $\xmid = \vec{x}_{(1)}$ and $\vec{x}_{(i)}$ is the location of the $i$th closest relay to $\vec{w}$. Then, by Lemma \ref{Lemma: Orientation Property}, $\sqrt{d^2 + \norm{\vec{x}_{(i)} - \vec{w}}^2} \leq \relayselect\paren{\vec{x}_{(i)}}$ for all $i \geq 1$. Further, $\sqrt{d^2 + \norm{\vec{x}_{(2)} - \vec{w}}^2} \leq \sqrt{d^2 + \norm{\vec{x}_{(i)} - \vec{w}}^2}$ for all $i \geq 2$. Hence, whenever $\relayselect\paren{\xmid} \leq \sqrt{d^2 + \norm{\vec{x}_{(2)} - \vec{w}}^2}$, we have $\relayselect\paren{\xmid} = \min_{\vec{x} \in \varphi} \relayselect\paren{\vec{x}}$.     
\end{IEEEproof}

For any realization of relay locations, this condition is easy to check since only location information of the closest and second closest relay nodes to the mid-point needs to be known.  
\vspace{0mm}
\section{Numerical Results} \label{Section: Numerical Results}
In our simulations, all distances 
are normalized to a {\it unit} distance. 
A circular network coverage area with radius $10$ [units] is considered. The path-loss exponent $\alpha$ is set to $4$. Results are obtained  by averaging over $10^6$ realizations of the random network topology and channel conditions.  To benchmark the optimum relay selection policy, we also simulate the network performance when the relay is chosen in a such way that it is closet to the mid-point, to the source or to the destination. 

\begin{figure}[!htb]
\centering
\includegraphics[width=0.7\textwidth]{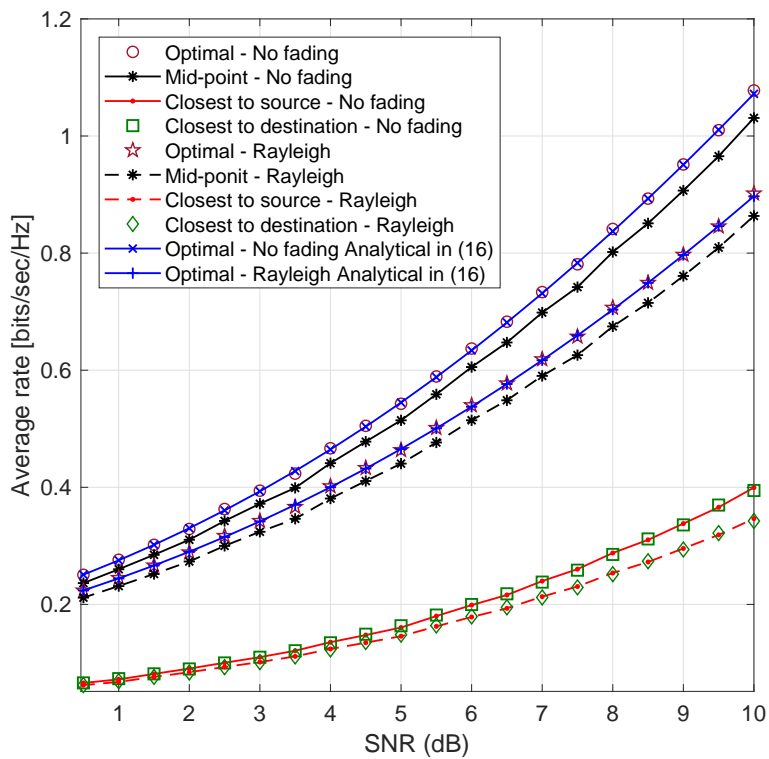}
\caption{Average rate vs \snr \,\,for different selection schemes.}\label{f_tpvssnr}
\end{figure}

\begin{figure}[!htb]
\centering
\includegraphics[width=0.7\textwidth]{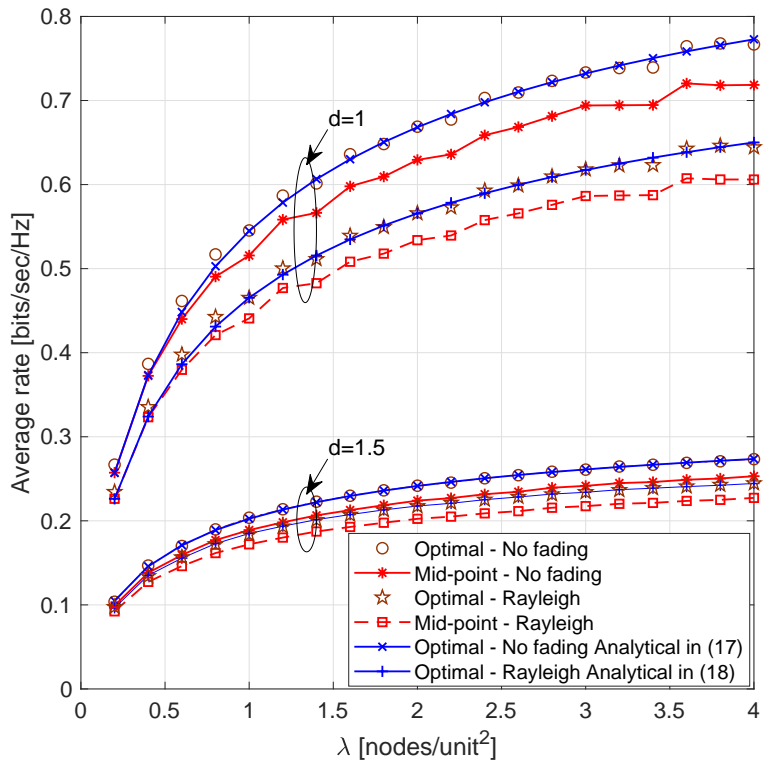}
\caption{Average rate vs $\lambda$ for \snr=5\,dB.}\label{f_tpvslam}
\end{figure}

\begin{figure}[!htb]
\centering
\includegraphics[width=0.7\textwidth]{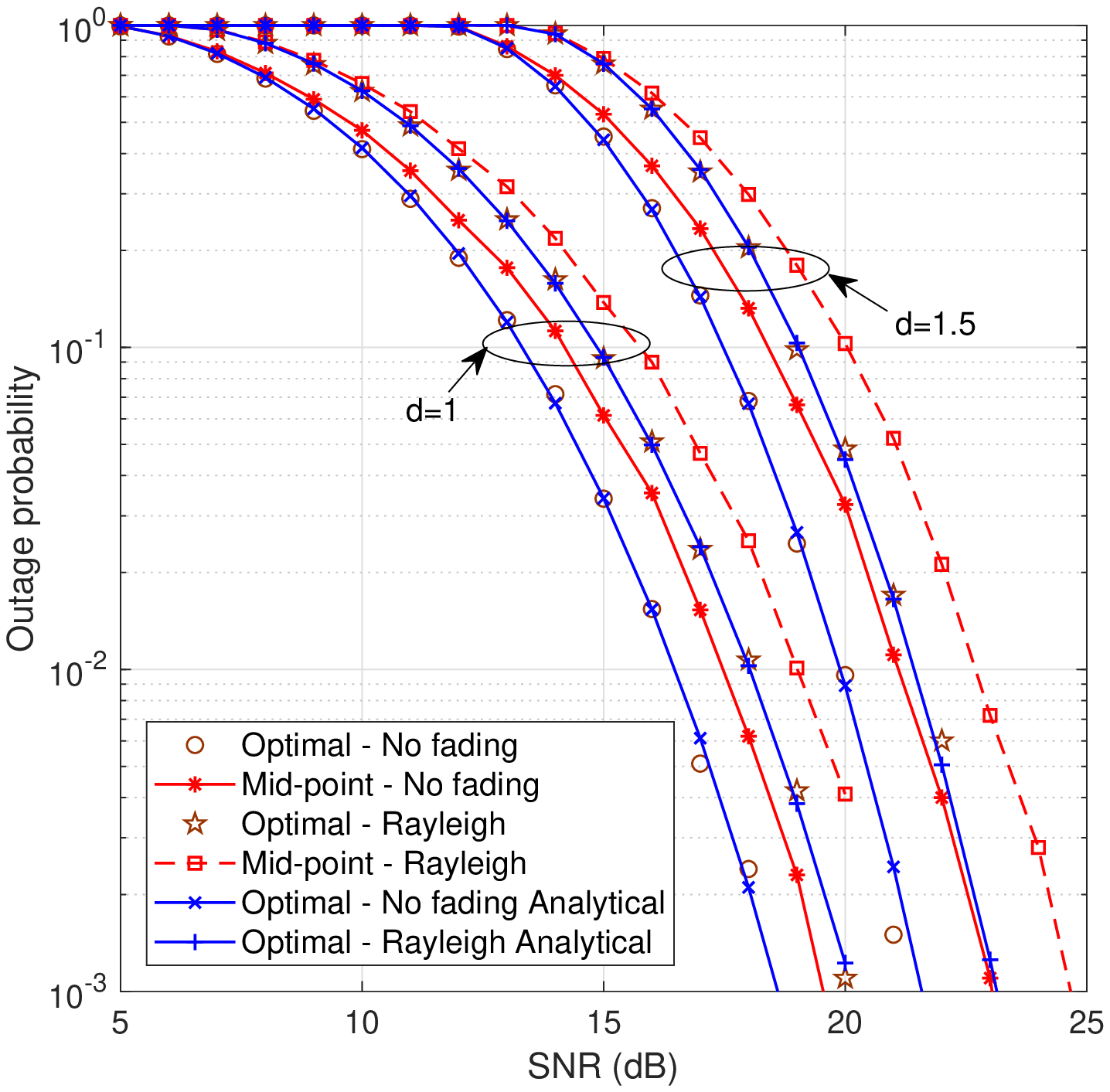}
\caption{Outage probability vs $\snr$ for $\lambda=1$ [nodes/unit$^2$].}\label{f_outvssnr}
\end{figure}

Fig.~\ref{f_tpvssnr} shows the average rate curves as a function of $\snr$ for all four relay selection schemes with and without fading when $\lambda=1$ [nodes/unit$^2$] and $d=1$ [unit]. This figure verifies the average rate expressions in \eqref{e_avgtp}.   
Several additional observations are: 
i) Optimum relay selection outperforms all other relay selection schemes; ii) closet-to-source and -destination schemes have similar but the worst performance; iii) no-fading scenario outperforms the fading case as expected; and iv)  there is a non-negligible performance gap between the optimum and mid-point relay selection schemes. For example,  the optimum relay selection provides around $8$\% improvement over the mid-point selection at $\snr=5$ dB. This may be especially significant for mmWave communications. 

Fig.~\ref{f_tpvslam} shows the average rate as a function of $\lambda$ for the optimum and mid-point relay selections with and without fading when $d=1$ [unit] and $d=1.5$ [units]. While we notice similar performance variations for the optimum and mid-point relay selection policies in Fig.~\ref{f_tpvslam}, 
we also observe that the average rate monotonically increases with $\lambda$ and reaches to $\frac{1}{2}\log_2\left(1+\frac{\snr }{d^\alpha}\right)$ and $\frac{1}{2\ln2} e^{\frac{d^\alpha}{\snr}} \text{E}_1\left(\frac{d^\alpha}{\snr}\right)$ for the no-fading and Rayleigh fading cases, respectively. 
For $d=1$ [unit], these asymptotic values are $1.03$ [bits/sec/Hz] for the no-fading case and $0.35$ [bits/sec/Hz] for the Rayleigh fading case. Further, we can see that there is a significant performance gap when the distance increases. For example, we lose $0.43$ [bits/sec/Hz] and $0.35$ [bits/sec/Hz] at $\lambda=2$ for the no-fading and Rayleigh fading cases, respectively, for the optimal selection. 

Fig.~\ref{f_outvssnr} shows the outage probability as a function of $\snr$ for the optimum and mid-point relay selections with and without fading when $\rho = 1$ [bits/sec/Hz], $d=1$ [unit] and $d=1.5$ [units]. This figure verifies the outage probability expressions in \eqref{e_pout_nofad} and \eqref{e_pout_ray}. The optimum selection outperforms the mid-point one for all cases. In particular, the outage probability of $0.01$ is achieved at $\snr\approx 16.5$ dB for the optimum selection and at $\snr\approx 17.5$ dB for the mid-point selection, which is a $1$ dB power gain. For the same outage probability with the optimum selection, we lose around $3.5$ dB by increasing the source-destination distance from $d=1$ [unit] to $d=1.5$ [units].   
\vspace{-0mm}
\section{Conclusions}
In this paper, we have considered a relay-aided wireless network with a single source-destination pair and spatially deployed decode-and-forward relays. For the optimum location-based relay selection policy, 
we have derived the distribution for the optimum relay distance in closed-form. Subsequently, we have evaluated the average rate and outage probability for the no-fading and Rayleigh fading channels by means of the derived optimum relay distance distribution.  To benchmark the optimum relay selection scheme, we have analyzed the mid-point relay selection policy and obtained a sufficient condition for its optimality. The numerical results have indicated that the optimum policy outperforms other existing policies in the literature. Further, the performance improvement becomes significant especially for emerging applications requiring wider bandwidths such as mmWave communications. 
\appendices
\section{Proof of Theorem~\ref{Theorem: Optimal dis distribution}}\label{Appendix: optimal dis distribution Proof}

We first divide the relay locations into two types:
\begin{eqnarray}
\Phi_{\rm right} = \Phi \bigcap \R^2_{\rm right} \,\, \text{and}\,\,\Phi_{\rm left} = \Phi \bigcap \R^2_{\rm left}, 
\end{eqnarray}
where $\R^2_{\rm right} = \brparen{\paren{x_1, x_2}^\top \in \R^2: x_1 \geq 0}$ and $\R^2_{\rm left} = \brparen{\paren{x_1, x_2}^\top \in \R^2: x_1 < 0}$. That is, the relays in $\Phi_{\rm right}$ are closer to the destination node, whereas the ones in $\Phi_{\rm left}$ are closer to the source node. Then, we define 
\begin{eqnarray}
\Gamma_{\rm opt}^{\rm right} \defeq \min_{\vec{X} \in \Phi_{\rm right}} \relayselect\paren{\vec{X}} \,\, \text{and}\,\, \Gamma_{\rm opt}^{\rm left} \defeq \min_{\vec{X} \in \Phi_{\rm left}} \relayselect\paren{\vec{X}}. 
\end{eqnarray}
Due to stationarity of HPPPs and symmetry of the problem, $\Gamma_{\rm opt}^{\rm right}$ and $\Gamma_{\rm opt}^{\rm left}$ are identically distributed random variables. Further, they are also independent due to complete randomness property of PPPs \cite{Kingman93}. Hence, it will be enough to obtain the cdf of $\Gamma_{\rm opt}^{\rm right}$ to prove Theorem \ref{Theorem: Optimal dis distribution} since $\Gamma_{\rm opt} = \min\brparen{\Gamma_{\rm opt}^{\rm right}, \Gamma_{\rm opt}^{\rm left}}$. More specifically,   
$F_{\Gamma_{\rm opt}}\paren{\gamma} = 1 - \paren{1 - F_{\Gamma_{\rm opt}^{\rm right}}\paren{\gamma}}^2$. 
To obtain $F_{\Gamma_{\rm opt}^{\rm right}}\paren{\gamma}$, we further define $\Phi_{{\rm right}, \tau} \defeq \Phi_{\rm right} \cap \mathcal{B}\paren{\vec{0}, \tau}$ and $\Gamma_{{\rm opt}, \tau}^{\rm right} \defeq \min_{\vec{X} \in \Phi_{{\rm right}, \tau}}\relayselect\paren{\vec{X}}$, where $\mathcal{B}\paren{\vec{0}, \tau}$ is the disc centered around the origin $\vec{0}$ and having radius $\tau$. We note that $\Gamma_{{\rm opt}, \tau}^{\rm right}$ converges almost surely to $\Gamma_{\rm opt}^{\rm right}$ as $\tau$ tends to infinity. Thus, the cdf of $\Gamma_{{\rm opt}, \tau}^{\rm right}$ will also converge to the cdf of $\Gamma_{\rm opt}^{\rm right}$ pointwise as $\tau$ tends to infinity \cite{Billingsley95}. We will derive the cdf of $\Gamma_{\rm opt}^{\rm right}$ by first obtaining the cdf of $\Gamma_{{\rm opt}, \tau}^{\rm right}$ and then taking the limit $\tau \ra \infty$.    

Let $N$ be the number of relays in $\Phi_{{\rm right}, \tau}$. Given the event $\brparen{N=n}$ for $n\geq 1$, all the relays in $\Phi_{{\rm right}, \tau}$ will be uniformly distributed over the half-disc centered at $\vec{0}$, having radius $\tau$ and containing only those points of $\R^2$ with non-negative first coordinates. Let $\vec{U}$ be such a uniformly distributed random relay location. Let also $\Gamma = \relayselect\paren{\vec{U}}$, which is equal to the distance between $\vec{U}$ and the source node. $\Gamma$ can be written as 
$\Gamma = \sqrt{\norm{\vec{U}}^2 + 2d\norm{\vec{U}} \cos\Theta + d^2}$
by using the law of cosines, where $\Theta$ is the angle between the positive $x$-axis and the line segment connecting $\vec{0}$ and $\vec{U}$. $\Theta$ is uniformly distributed over $\sqparen{-\frac{\pi}{2}, \frac{\pi}{2}}$, and independent of $\norm{\vec{U}}$ due to the uniformly distributed nature of $\vec{U}$. Hence, the conditional cdf of $\Gamma$ given $\brparen{\Theta = \theta}$ can be expressed as 
\begin{eqnarray}
F_{\Gamma | \Theta}\paren{\gamma | \theta} &=& \PR{\Gamma^2 \leq \gamma^2 | \Theta = \theta} \nonumber \\
&=& \PR{\norm{\vec{U}} \leq \sqrt{\gamma^2 - d^2\sin^2\theta} - d\cos\theta} \nonumber
\end{eqnarray}
for $d \leq \gamma \leq \sqrt{\tau^2 + 2d\tau\cos\theta + d^2}$, where the last equation follows from the fact that the convex quadratic function $f(x) = x^2 + 2d \cos\theta x + d^2-\gamma^2$ has only one positive root at $\sqrt{\gamma^2 - d^2\sin^2\theta} - d\cos\theta$ when $\gamma \geq d$. For $\gamma < d$, we have $F_{\Gamma | \Theta}\paren{\gamma | \theta} = 0$ since $\Gamma$ is always greater than or equal to $d$. For $\gamma > \sqrt{\tau^2 + 2d\tau\cos\theta + d^2}$, we have $F_{\Gamma | \Theta}\paren{\gamma | \theta} = 1$ since $\Gamma$ is always smaller than or equal to $\sqrt{\tau^2 + 2d\tau\cos\theta + d^2}$ when $\Theta = \theta$. As a result, using the cdf of $\norm{\vec{U}}$, which is equal to $F_{\norm{\vec{U}}}(u) = \frac{u^2}{\tau^2}$, we have 
\begin{equation}\label{Eqn: Conditional Distance CDF}
F_{\Gamma | \Theta}(\gamma | \theta)
=\left\{
\begin{array}{ll}
0 \qquad\qquad\mbox{ if } \gamma < d\\
\frac{\left(\sqrt{\gamma^2-d^2 \sin^2\theta}-d\cos\theta\right)^2}{\tau^2} \\
 \quad\mbox{ if } d \leq \gamma \leq \sqrt{\tau^2 + 2d\tau\cos\theta + d^2} \\   
1 \qquad\quad\mbox{ if } \gamma > \sqrt{\tau^2 + 2d \tau \cos\theta + d^2}
\end{array}\right.
\end{equation}
\begin{figure*}
{\small
\begin{equation}\label{Eqn: Distance Distribution}
F_{\Gamma}(\gamma)
=\left\{
\begin{array}{ll}
0& \mbox{ if } \gamma < d \\
\frac{2}{\pi \tau^2}\left(\gamma^2 \arcsec\left(\frac{\gamma}{d}\right)- d \sqrt{\gamma^2-d^2}\right)& \mbox{ if } d\leq \gamma \leq \sqrt{\tau^2+d^2} \\
\frac{2\gamma^2}{\pi \tau^2}\left(\arcsec\left(-\frac{2 d \tau}{\tau^2+d^2-\gamma^2}\right) - \arctan\left(\frac{\sqrt{4 d^2 \tau^2-\left(\tau^2+d^2-\gamma^2\right)^2}}{\tau^2-d^2+\gamma^2}\right) \right)&  \\
\hspace{0.1cm} - \frac{2 \arccsc\left(\frac{2 d \tau}{\tau^2+d^2-\gamma^2}\right)}{\pi}  -\frac{\sqrt{(\tau-d+\gamma)(\tau+d-\gamma)(d-\tau+\gamma)(\tau+d+\gamma)}}{\pi \tau^2}& \mbox{ if } \sqrt{\tau^2+d^2} < \gamma \leq \tau+d \\ 
1& \mbox{ if } \gamma > \tau+d
\end{array}.\right.
\end{equation}}
\hrule
\end{figure*}
We will obtain $F_{\Gamma}(\gamma)$ by averaging \eqref{Eqn: Conditional Distance CDF} over $\Theta$.
To this end, we need to consider four cases separately. If $\gamma < d$, then $F_{\Gamma | \Theta}(\gamma | \theta) = 0$ for all $\theta \in \sqparen{-\frac{\pi}{2}, \frac{\pi}{2}}$. Hence, $F_{\Gamma}(\gamma) = 0$ when $\gamma < d$. If $d \leq \gamma \leq \sqrt{\tau^2 + d^2}$, the second condition in \eqref{Eqn: Conditional Distance CDF} is always satisfied, and we have 
\begin{eqnarray*}
F_{\Gamma}(\gamma) = \frac{2}{\pi \tau^2}\int_0^{\frac{\pi}{2}} \paren{\sqrt{\gamma^2-d^2\sin^2\theta} - d\cos\theta}^2d\theta
\end{eqnarray*}
for this range of $\gamma$. If $\sqrt{\tau^2 + d^2} < \gamma \leq \tau+d$, the second and third conditions in \eqref{Eqn: Conditional Distance CDF} are satisfied for $\theta \in \sqparen{-\theta^\star, \theta^\star}$ and $\theta \in \parenro{-\frac{\pi}{2}, -\theta^\star} \cup \parenlo{\theta^\star, \frac{\pi}{2}}$, respectively, where $\theta^\star = \arccos\paren{\frac{\gamma^2 - \tau^2 - d^2}{2d\tau}}$. Thus, we have  
\begin{eqnarray*}
F_{\Gamma}(\gamma) = \frac{2}{\pi \tau^2}\hspace{-0.1cm}\int_0^{\theta^\star} \hspace{-0.2cm}\paren{\sqrt{\gamma^2-d^2\sin^2\theta} - d\cos\theta}^2\hspace{-0.1cm}d\theta + 1 -\frac{2\theta^\star}{\pi}  
\end{eqnarray*}
for this range of $\gamma$. Finally, if $\gamma > \tau+d$, then $F_{\Gamma | \Theta}(\gamma | \theta) = 1$ for all $\theta \in \sqparen{-\frac{\pi}{2}, \frac{\pi}{2}}$, and therefore $F_{\Gamma}(\gamma) = 1$ if $\gamma > \tau + d$. Combining all four cases and evaluating the integrals, we obtain $F_{\Gamma}(\gamma)$ as in \eqref{Eqn: Distance Distribution}. 
We use $F_{\Gamma}(\gamma)$ and obtain 
\begin{eqnarray}
F_{\Gamma_{{\rm opt}, \tau}^{\rm right}}(\gamma) &=& \sum_{n=0}^\infty \paren{1-\paren{1-F_{\Gamma}\paren{\gamma}}^n} \PR{N=n} \nonumber \\
&=& 1 - \sum_{n=0}^\infty \paren{1-F_{\Gamma}\paren{\gamma}}^n \frac{\paren{\frac{\lambda\pi \tau^2}{2}}^n \e{-\frac{\lambda \pi \tau^2}{2}}}{n!} \nonumber \\
&=& 1 - \e{-\frac{\lambda \pi \tau^2}{2}F_{\Gamma}\paren{\gamma}}. \label{Eqn: Distance Distribution 2} 
\end{eqnarray}
As stated earlier, $F_{\Gamma_{{\rm opt}}^{\rm right}}(\gamma) = \lim_{\tau \ra \infty} F_{\Gamma_{{\rm opt}, \tau}^{\rm right}}(\gamma)$ since $\Gamma_{{\rm opt}, \tau}^{\rm right}$ converges to $\Gamma_{{\rm opt}}^{\rm right}$ almost surely. Thus, by using \eqref{Eqn: Distance Distribution 2}, we have  
\begin{equation} \label{Eqn: Distance Distribution 3}
\begin{split}
F_{\Gamma_{{\rm opt}}^{\rm right}}(\gamma) &= \lim_{\tau \ra \infty} F_{\Gamma_{{\rm opt}, \tau}^{\rm right}}(\gamma) 
= 1 - \lim_{\tau \ra \infty} \e{-\frac{\lambda \pi \tau^2}{2}F_{\Gamma}\paren{\gamma}} \\
&= \left\{
\begin{array}{ll}
0& 
\gamma < d \\
1-\e{\lambda \paren{d \sqrt{\gamma^2-d^2} - \gamma^2 \arcsec\left(\frac{\gamma}{d}\right)}}& 
\gamma \geq d 
\end{array}.\right.
\end{split}
\end{equation}
Finally, using \eqref{Eqn: Distance Distribution 3} 
and the identity $F_{\Gamma_{\rm opt}}\paren{\gamma} = 1 - \paren{1 - F_{\Gamma_{\rm opt}^{\rm right}}\paren{\gamma}}^2$, we conclude the proof. 
\vspace{-0mm}
\section{Proof of Lemma \ref{Lemma: Orientation Property}}\label{Appendix: Orientation Property Proof}
\vspace{0mm}
Assume $\vec{y}$ belongs to the half-space closer to $\xd$ and consider the decomposition $\vec{y} = \vec{w} + \vec{z}$ for some $\vec{z} \in \R^2$. Then, we have 
$\paren{\xs - \xd}^\top \vec{z} \leq 0$ and 
\begin{eqnarray}
\relayselect\paren{\vec{y}} &= \norm{\xs - \vec{y}} 
=\sqrt{\norm{\xs - \vec{w}}^2 - \paren{\xs - \xd}^\top \vec{z} + \norm{\vec{z}}^2} \nonumber \\
&\geq \sqrt{\norm{\xs - \vec{w}}^2 + \norm{\vec{y} - \vec{w}}^2}. \nonumber 
\end{eqnarray}
Similarly, for any $\vec{x} \in \mathcal{H} \cap \mathcal{C}$, we have $\relayselect\paren{\vec{x}} = \sqrt{\norm{\xs - \vec{w}}^2 + \norm{\vec{y} - \vec{w}}^2}$. Hence, $\relayselect\paren{\vec{x}} \leq \relayselect\paren{\vec{y}}$. The arguments for when $\vec{y}$ belongs to the half-space closer to $\xs$ are the same.   


\end{document}